\documentclass[10pt,journal,compsoc]{IEEEtran}
\usepackage[utf8]{inputenc}

\usepackage[english]{babel}
\usepackage{amsmath, amssymb, amsthm, amscd,color}

\usepackage[pdftex]{graphicx}

\usepackage{cancel}
\usepackage{cite}
\usepackage{alltt}

\usepackage{enumerate}
\usepackage{cancel}
\newtheorem{theorem}{Theorem}[section]
\newtheorem{definition}[theorem]{Definition}
\newtheorem{example}[theorem]{Example}

\newtheorem{remark}[theorem]{Remark}
\newtheorem{lemma}[theorem]{Lemma}
\newtheorem{corollary}[theorem]{Corollary}
\newtheorem{proposition}[theorem]{Proposition}

\newenvironment{keywords}{\begin{quote}{\bf Keywords:}}{\end{quote}}
\newenvironment{msccodes}{\begin{quote}{\bf MSC classification codes:}}{\end{quote}}

\def\fq{\mathbb{F}_{q}}
\def\fqs{\mathbb{F}_{q^2}}

\def\Pla{\mathcal{P}}
\def\cF{\mathcal{F}}

\def\cC{\mathcal{C}}
\def\cF{\mathcal{F}}

\def\cH{\mathcal{H}}
\def\cL{\mathcal{L}}

\def\N{\mathbb{N}}

\def\cF{\mathcal{F}}
\DeclareMathOperator\supp{supp}

\newcommand\mut[1]{\ignorespaces}

\newcommand\graphhsmallone{}
\newcommand\graphntsmallone{}
\newcommand\graphhsmalltwo{}
\newcommand\graphntsmalltwo{}
\newcommand\graphh{}
\newcommand\graphnt{}
\newcommand\graphk{}

\input{graphsIEEE}

\title{The Isometry-Dual Property in Flags of \\Two-Point Algebraic Geometry Codes}

\date{\today}

\begin{document}

\author{Maria Bras-Amorós, Alonso S. Castellanos, Luciane Quoos
  \thanks{M. Bras-Amor\'os is with the Department of Computer Engineering and Mathematics, Universitat Rovira i Virgili, Catalonia (e-mail: maria.bras@urv.cat).}
  \thanks{L.~Quoos is with the Instituto de Matem\'atica at the Universidade Federal do Rio de Janeiro, Brazil (e-mail: luciane@im.ufrj.br).}
  \thanks{Alonso~S.~Castellanos is with the Faculdade de Matem\'atica at the Universidade Federal de Uberl\^andia, Brazil (e-mail: alonso.castellanos@ufu.br).} 
  \thanks{The authors would like to thank the funding from FAPERJ \#88881.360642/2019-01 that made their collaboration possible.
They would also like to thank Prof. Kwankyu Lee, from Chosun University, Korea, for his help in the usage of {\sc Sage}.
The first author was partially supported by Consell Català de Recerca i Innovació (2017 SGR 00705) and Ministerio de Ciencia, Innovación y Universidades (TIN2016-80250-R, RTI2018-095094-B-C21).
The second author was partially supported by Fundação de Amparo à Pesquisa do Estado de Minas Gerais (FAPEMIG APQ- 00696-18). The third author would like to thank CNPq  302727/2019-1 and FAPERJ 260003/002364/2021. }
    \thanks{Copyright (c) 2017 IEEE. Personal use of this material is permitted.  However, permission to use this material for any other purposes must be obtained from the IEEE by sending a request to pubs-permissions@ieee.org.}
}
%
\maketitle

\begin{abstract}
A flag of codes $C_0 \subsetneq C_1 \subsetneq \cdots \subsetneq C_s \subseteq \fq^n$ is said to satisfy the {\it isometry-dual property} if  there exists ${\bf x}\in (\mathbb{F}_q^*)^n$ such that the  code $C_i$ is {\bf x}-isometric to  the dual code $C_{s-i}^\perp$ for all $i=0,\ldots, s$. 
For $P$ and $Q$ rational places in a function field $\cF$, we investigate the existence of isometry-dual flags of codes in the families of two-point algebraic geometry codes $$C_\mathcal L(D, a_0P+bQ)\subsetneq
C_\mathcal L(D, a_1P+bQ)\subsetneq \dots \subsetneq C_\mathcal L(D, a_sP+bQ),$$
where the divisor $D$ is the sum of pairwise different rational places of $\cF$ and $P, Q$ are not in $\supp(D)$. We characterize those sequences in terms of $b$ for general function fields. 
We then apply the result to the broad class of Kummer extensions $\cF$ defined by affine equations of the form $y^m=f(x)$, for $f(x)$ a separable polynomial of degree $r$, where $\gcd(r, m)=1$.
For $P$ the rational place at infinity and $Q$ the rational place associated to one of the roots of $f(x)$,
and for 
$D$ an $Aut(\cF/\fq)$-invariant sum of rational places of $\cF$, such that $P, Q \notin \supp D$,
it is shown that the flag of two-point algebraic geometry codes has the isometry-dual property if and only if $m$ divides $2b+1$. 
At the end we illustrate our results by applying them to two-point codes over several well know function fields.
\end{abstract}


\begin{msccodes}
  14G50,        
  11T71,   	
  94B27,   	
  14Q05   	
\end{msccodes}

\section{Introduction}

A linear code $C\subseteq\fq^n$ is a $\fq$-linear subspace of $\fq^n$. The dual code $C^\perp$ of $C$ is defined as the orthogonal complement of $C$ in $\fq^n$ with the standard inner product of $\fq^n$.
The explicit determination of dual codes is specially important in the detection and correction of errors.
A code is {\em self-dual} if $C=C^\perp$ and such codes have been investigated in \cite{MST1,MP1993} and recently in \cite{IEEEselfdual, S2021}.
This is a restrictive condition and can be relaxed to the condition $C \subseteq C^\perp$, defining the so-called {\em self-orthogonal} codes. In \cite{BMZ2021, MTF2016} and \cite{LP2017} we find applications of self-orthogonal algebraic geometry codes to the construction of quantum codes using the {\it CSS} construction introduced in \cite{KM2008}. 

A  {\it flag} of linear codes $C_0 \subsetneq C_1 \subsetneq \cdots \subsetneq C_s$
is said to have the {\it isometry-dual property} (first introduced in \cite{GMRT}) if there exists ${\bf x}\in (\mathbb{F}_q^*)^n$ such that $C_i$ is {\bf x}-isometric to $C_{s-i}^\bot$ for all $i=0,\ldots, s$, that is, $C_i={\bf x} \cdot C_{s-i}^\perp$.

The theory of algebraic geometry codes attracted a lot of attention in the last decades since the construction of linear codes from algebraic curves by Goppa in 1982 \cite{Goppa}. 
Let  $\cF/\fq$ be a function field. Consider $D=P_1+\dots+P_n$ a divisor given by the sum of pairwise distinct rational places on $\cF$,  and $G$ a divisor such that $P_i$ is not in the support of $G$ for $i = 1,\dots ,n$. The algebraic geometry code $C_{\mathcal{L}}(D, G)$ (abbreviated by AG code) is defined by
 $$C_{\mathcal{L}}(D, G)=\{ (f(P_1),\ldots, f(P_n))\mid f\in \mathcal{L}(G)\} \subseteq \fq^n,$$
where $\cL(G)$ denotes the Riemann-Roch space associated to the divisor $G$. 
These codes are called {\it one-point} or {\it two-point codes} provided the divisor $G$ is supported on a single rational place $P$ (that is, $G=aP$), or on two different rational places $P, Q$ (that is, $G=aP+bQ$), respectively. We notice that a point for us is the same as a rational place. Determining or even improving the parameters of one or two-point AG codes has been a major subject of research, see for example \cite{HommaKim:Thecomplete, HK2001, KimLee, XC2002}, and \cite{CT2005}.

Natural flags of one-point codes are obtained by varying $a$ in the divisor $G=aP$. In \cite{GMRT,BDH} one can find an analysis of flags of one-point AG codes satisfying the isometry-dual property.  Explicit constructions are carried on over the projective line, Hermitian, Suzuki, Ree, and the Klein function fields. In \cite{BLV} there is a characterization of isometry-dual flags of one-point codes in terms of sparse ideals of numerical semigroups.

In this work we investigate the isometry-dual property in flags of two-point algebraic geometry codes.
Given a fixed $b \in \N$ and an increasing sequence $0\leq a_0\leq \dots \leq a_s$ we study the flag
\begin{equation*}\label{eq:seqcodes}
\begin{split}
C_\mathcal L(D, a_0P+bQ)&\subsetneq
C_\mathcal L(D, a_1P+bQ)\subsetneq\\ 
&\dots \subsetneq C_\mathcal L(D, a_sP+bQ)
\end{split}
\end{equation*} 
An important role in this investigation is played by the sets 
$$H_b=\{a\geq 0 \, :\,  \ell(aP+bQ)\neq \ell((a-1)P+bQ) \}$$
and 
$$H_b^*=\{a\geq 0 \, :\, C_\mathcal L(D, aP+bQ)\neq C_\mathcal L(D, (a-1)P+bQ) \}$$
that have a close connection with the Weierstrass semigroups at one rational place $H(P)$, and two rational places $H(P, Q)$. In Theorem \ref{t:isodual} this connection allows to provide an arithmetic condition on $b$ to decide when the flag \eqref{eq:seqcodes} has the isometry-dual property. 

From a coding theory point of view, 
it is natural to consider function fields with many rational places.
Indeed,
if $n$, $k$, $d$ are, respectively, the length, dimension, and minimum distance of an algebraic geometry code over a function field of genus $g$, the rate $(k+d)/n$ satisfies the relation
$$1+(1-g)/n \leq (k+d)/n \leq 1+1/n.$$ 
  These inequalities depend on the length $n$ of the code with respect to the genus $g$ of the underlying function field.
  Codes with a large rate $(k+d)/n$ are considered the best ones.
Thus, function fields with many rational places, that is, function fields over $\fq$ such that the number of rational places is close (or equal) to the Hasse-Weil upper bound  $q+1+2g\sqrt{q}$ have been intensively studied. A function field attaining the Hasse-Weil upper bound is called {\it  a maximal function field}.
Several explicit constructions of AG codes over well known families of maximal or with many rational places function fields can be found in the literature, see \cite{XC2002, MTT, Geil, KimLee, G2004}, and \cite{AM}. In these papers a variety of subjects are investigated, such as the computation or improvement of parameters.  It turns out that numerous examples of function fields with many rational places can be described by an affine equation of the form $y^m=f(x)$ for some polynomial $f(x) \in \fq[x]$. Important examples of maximal function fields of such form are the Hermitian function field   \cite{Gretchen}, the Giullietti-Korchmáros function field  \cite{GK2009}, and the generalized Hermitian function field described in \cite{KKO,GV1986}. Other examples of such function fields with  many rational places are given by the norm-trace function field \cite{Geil, MTT}, and the ones as the described in \cite{GQ2001}. Function fields defined by equations of the form $y^m=f(x)$ are also known as Kummer extensions, see \cite{Stichtenoth} for a formal definition. The study of codes on one or more points over certain Kummer extensions can be found in  \cite{CMQ, Geil, MTT, BQZ2016, SH2017}.

Using a result in \cite{Maharaj} we are able to investigate the isometry-dual property in the large class of algebraic Kummer extensions given by $y^m=f(x)$, where $f(x)$ is a separable polynomial of degree $r$, with $r, m$ coprime and 
$2\leq r\leq m-1$. These function fields have only one place at infinity, denoted by $P_\infty$. Fixing a second place $Q$ associated to a root of $f(x)$, in Theorem \ref{t:isodualKummer} we show that the flag \eqref{eq:seqcodes} satisfies the isometry-dual condition if and only if $m$ divides $2b+1$. 

The article is organized as follows. In Section 2, we present preliminary results on function fields, codes and Weierstrass semigroups. In Section 3, we investigate the dimension of certain Reimann-Roch spaces and codes associated to these spaces. In Theorem \ref{t:isodual} in Section 4, we present our main result concerning the isometry-dual property on flags of AG two-point codes. In Section 5, we present the case of two-point codes and the isometry-dual property over the Hermitian function field.
In Section 6, we present results for Kummer extensions in Theorems \ref{t:kummer} and \ref{t:isodualKummer}, and provide explicit results for the norm-trace function field and for the generalized Hermitian function field defined by the equation $y^{q^\ell+1}=x^q+x$ over $\mathbb{F}_{q^{2\ell}}$.
We implemented several examples in {\tt magma} and {\tt sage} to illustrate all the results presented in each section.

\section{Preliminaries}

In this section we state the notation and some preliminary results on function fields, algebraic geometry codes, and Weierstrass semigroups.

Let  $\cF/\fq$ be a function field.
We denote by $\Pla(\cF)$ the set of places of $\cF$, and the places of degree one over $\fq$ are called  {\it rational places}.
A divisor is a formal finite sum of places in $\Pla(\cF)$,
and the degree of a divisor $D=\sum_{P\in\Pla(\cF)}n_PP$ is the sum $\deg(D)=\sum_{P\in \Pla(\cF)}n_P\deg(P).$
For a function $z$ in $\cF$, $(z)$, $(z)_0$, and $(z)_\infty$ stand for its divisor, zero divisor, and pole divisor, respectively. 

Given a divisor $D$, the Riemann-Roch vector space associated to $D$ is defined by
$
\cL(D)=\{z\in \cF\,:\, (z)\ge -D\}\cup \{0\}.
$
We denote by $\ell(D)$ the dimension of $\cL(D)$ as a vector space over $\fq$.
Two divisors $D_1$ and $D_2$ are said to be equivalent if there exists a function $z \in \cF$ such that $D_1=D_2 + (z)$ and we write $D_1 \sim D_2$.
In this case, $\cL(D_1)=z^{-1}\cL(D_2)$
and it follows that the Riemann–Roch spaces $\cL(D_1)$ and  $\cL(D_2)$ are isomorphic.

In the present work we illustrate our results on the broad class of Kummer extensions defined by affine equations of the form $y^m=f(x)$, for $f(x)$ a separable polynomial of degree $r$ with $\gcd(m, r)=1$ and $2\leq r \leq m-1$.
This class contains important examples such as the Hermitian, the norm-trace, and the generalized Hermitian function field defined by the equation $y^{q^\ell+1}=x^q+x$ over $\mathbb{F}_{q^{2\ell}}$. In this case, the dimension of Riemann-Roch spaces can be analyzed by means of the next theorem due to Maharaj \cite{Maharaj}
by decomposing them as a direct sum of Riemann–Roch spaces of divisors over the projective line. At first we need a definition. For any function field extension $K \subseteq E\subseteq \cF$, and for a divisor $D$ of $\cF$, define the restriction of $D=\sum_{P\in\Pla(\cF)}n_PP$ to $E$ as the divisor 
$$D_{|E}= \sum_{R\in \Pla(E)} min\,\bigg\{\bigg\lfloor\frac{n_P}{e(P|R)}\bigg\rfloor:  P\in\Pla(\cF)\mbox{ and }{P|R}\bigg\}\,R,$$ where $e(P|R)$ is the ramification index of $P$ over~$R$.
\begin{theorem}\cite[Theorem 2.2]{Maharaj}\label{ThMaharaj}
  Let $\cF/K(x)$ be a Kummer extension of degree $m$ defined by $y^m=f(x)$.
  Then for any divisor $D$ of $\cF$,  with $D$ invariant by the action of $Gal(\cF/K(x))$, we have that
$$ \mathcal{L}(D)= \bigoplus\limits_{t=0}^{m-1}  \mathcal{L}([D+(y^t)]_{|K(x)})\,y^t.$$
\end{theorem}

Let $D$ and $G$ be divisors on $\cF$ such that $D$ is the sum of $n$ distinct rational places on $\cF$  that are not in the support of $G$; say, $P_1, \ldots, P_n$. The linear algebraic geometry code $C_{\cL}(D, G)$ is defined by 
 $$C_\cL(D, G)=\{ (f(P_1),\ldots, f(P_n))\, :\,  f\, \in \mathcal{L}(G)\} \subseteq \fq^n.$$
Through all the paper $n$, $k$, and $d$ will stand for the length, dimension, and minimum distance of $C_\cL(D,G)$, respectively. 
The next result establishes the relationship between these parameters and the divisor $G$.

\begin{proposition}{\cite[Theorem 2.2.2, Corollary 2.2.3]{Stichtenoth}}\label{dimcode} For the code $C_\cL(D, G)$,
$$k=\ell(G)-\ell(G-D) \text{ and } d \geq n-\deg(G).$$
Moreover, 
\begin{enumerate}[(1)]
\item if $\deg (G) < n$, then $k=\ell(G) \geq \deg(G)+1-g$, and
\item if $2g-2<\deg(G) < n$, then $k=\ell(G)= \deg(G)+1-g$.
\end{enumerate}
\end{proposition}

Let $P$ and $ Q$  be two rational places in $\Pla(\cF)$ and let $\mathbb{N}=\{0, 1, \dots \}$ denote the set of natural numbers. We define the {\em Weierstrass semigroup} at one place $P$, and two places $P, Q$ as 
$$
\begin{array}{ll}
H(P)=\{a\in{\mathbb N} \, :  \, \exists \, f \, \in \cF \mbox{\ with\ } (f)_\infty=aP\}, \text{ and }\\
H(P, Q)=\{(a, b)\in{\mathbb N}^2 \, :  \, \exists \, f \, \in \cF \mbox{\ with\ } (f)_\infty=aP+bQ\}.
 \end{array}
 $$
 One can check that $H(P)$ and $H(P, Q)$ are, indeed, subsemigroups of ${\mathbb N}$ and ${\mathbb N}^2$, respectively.
It follows by definition that $a \in H(P)$ if and only if $\ell(aP)\neq \ell((a-1)P)$ and, in a similar way, $(a, b) \in H(P, Q)$ if and only if $\ell(aP+bQ)\neq \ell((a-1)P+bQ) \text{ or } \ell(aP+bQ)\neq \ell(aP+(b-1)Q)$.
The {\em gap} sets are the complement sets $G(P)={\mathbb N}\setminus H(P)$ and $G(P, Q)={\mathbb N}^2\setminus H(P, Q)$.
For the semigroup defined at one rational place $P$ the set $G(P)$ has $g$ elements \cite[Theorem 1.6.8]{Stichtenoth} and its maximum element is at most $2g-1$. 
For the semigroup at two places, in general the cardinality of the gap set $G(P, Q)$ depends on the fixed places, see \cite{HK2001} and Section 3 in \cite{BQZ2016}. 

In the case of two rational places $P$ and $Q$, the knowledge of the Weierstrass semigroup at each one of the places allows a full description of the Weierstrass semigroup $H(P, Q)$  as we describe below.  Let $\beta_{1} < \beta_{2} < \cdots < \beta_{g}$  and
 $\gamma_{1} < \gamma_{2} < \cdots < \gamma_{g}$ be the gap sequences at $P$ and $Q$, respectively. For each $i$, let $n_{\beta_i} = \min \{ \gamma \in \mathbb{N} \, :\, (\beta_i, \gamma ) \in H(P, Q) \}$. Then $\{ n_{\beta} \, :\,  \beta \in G(P) \} = G(Q)$ by~\cite[Lemma 2.6]{kim}.  So there exists a permutation $\sigma$ of the set $\{1,2, \ldots, g\}$ such that $n_{\beta_{i}} = \gamma_{\sigma(i)}$. The graph of the bijective map between $G(P)$ and $G(Q)$ is the set
\begin{align*}
\Gamma(P, Q)  &= \{ (\beta_{i}, n_{\beta_i}) \, :\,  i=1,2, \ldots , g\} \\
&= \{ (\beta_{i}, \gamma_{\sigma(i)}) \, :\,  i=1,2, \ldots , g \} \\
& \subseteq G(P) \times G(Q).
\end{align*}
Given $\Gamma (P, Q)$, we can compute $H(P, Q)$ in the following way. For two pairs $\mathbf{x} = (\beta_{1}, \gamma_{1})$ and $\mathbf{y} = (\beta_{2}, \gamma_{2}) \in \mathbb{N}^2$, the \textit{least upper bound} of $\mathbf{x}$ and $\mathbf{y}$ is defined as $\mathrm{lub}(\mathbf{x},\mathbf{y})= (\max\{\beta_{1}, \beta_{2}\}, \max\{\gamma_{1}, \gamma_{2}\})$.

The elements in a semigroup always satisfy the following property concerning the lub function.
\begin{lemma}\cite[Lemma 2.2]{kim}\label{lubsemigroup}
If ${\bf u}_1$ and ${\bf u}_2\in H(P, Q)$, then $\mathrm{lub}({\bf u}_1,{\bf u}_2)\in H(P, Q)\;.$
\end{lemma}
In general we have
\begin{lemma}{\cite[Lemma 2.3]{kim}}
\label{l:Gamma}
Let $P$ and $Q$ be two distinct rational places. The Weierstrass semigroup associated to $P$ and $Q$ is given by
\begin{align*}
H(P, Q) &= \{ \mathrm{lub} (\mathbf{x},\mathbf{y}) \, : \, \\
& \mathbf{x},\mathbf{y} \in \Gamma(P, Q) \cup (H(P) \times \{0\}) \cup (\{0\} \times H(Q)) \}.
\end{align*}
\end{lemma}

\section{Dimension of two-point Riemann-Roch spaces and dimension of two-point codes}

Let $P, Q$ be two different rational places on the function field $\cF/\fq$ of genus $g$.
For $b$ in ${\mathbb N}$ define the set 
$$H_b=\{a\geq 0 \, :\,  \ell(aP+bQ)\neq \ell((a-1)P+bQ) \}.$$

The next lemma relates the set $H_b$ with the sets $H(P), H(Q),$ and $ H(P,Q)$.

\begin{lemma}\label{hinhb}
\begin{enumerate}[(1)]
\item For $b\in \mathbb N$, $\{a \geq 0\, :\, (a, b)\in H(P,Q) \}\subseteq H_b$, and if $b\in H(Q)$ then
$H_b=\{a \geq 0\, :\, (a, b)\in H(P,Q) \}.$
\item
  If  $b\in H(Q)$, then $H(P)\subseteq \{a \geq 0\, :\, (a, b)\in H(P,Q) \}$.
\item For $b\in\mathbb N$,
  if $a \geq 2g-b$ then  $a\in H_b$.

\end{enumerate}
\end{lemma}
    \begin{proof}
\begin{enumerate}[(1)]
\item The first inclusion follows by definition. 
Suppose that  $b\in H(Q)$ and let $a\in H_b$. Then there exists a rational function $f\in
 \cL(aP+bQ)\setminus\cL((a-1)P+bQ)$. So, the pole divisor of
        $f$ is $(f)_\infty=aP+rQ$ with $r\leq b$. This implies that
        $(a, r)\in H(P,Q)$ and as $(0, b)\in H(P, Q)$ then from
        Lemma \ref{lubsemigroup} we have  $(a, b)\in H(P,
        Q)$.  
\item Since $H(P)\times H(Q) \subseteq H(P, Q)$, we get 
  $H(P)\subseteq\{a \geq 0\, :\, (a, b)\in H(P,Q) \}.$

\item  If $a\geq 2g-b$, then $\deg(aP+bQ)\geq 2g$ and we get $\ell(aP+bQ)= \ell((a-1)P+bQ)+1$.
 
\end{enumerate}
\end{proof}

    \begin{example}
Consider the Hermitian function field over ${\mathbb F}_4$. In Figure~\ref{fig:Hermitefunctiondimensionnongapstwo} we depicted the sets $H_b$, $H(P,Q)$, and $H(Q)$, so that all the items in Lemma~\ref{hinhb} can be checked.
Details of this function field will be explicited in Section~\ref{s:H}.
The set $\Gamma(P,Q)$ was computed using the results in \cite[Theorem 3.4]{CT2018}.
Notice that in this case the unique gap of $H(Q)$ is $b=1$.
We can check that for any $b$ it holds $\{a \geq 0\, :\, (a, b)\in H(P,Q) \}\subseteq H_b$. Furthermore, for $b$ outside $1$ the inequality is, indeed, an equality and it holds
$H(P)\subseteq \{a \geq 0\, :\, (a, b)\in H(P,Q) \}$. But this is not true for $b=1$.

Similarly, consider the norm-trace function field over ${\mathbb F}_8$. In Figure~\ref{fig:NormTracefunctiondimensionnongapstwo} we depicted the sets $H_b$, $H(P,Q)$ and $H(Q)$.
Details of this function field will be explicited in Section~\ref{s:K}.
Again, the set $\Gamma(P,Q)$ was computed using the results in \cite[Theorem 3.4]{CT2018}.
In this case the genus is $g=9$.
We can check that for any $b$ it holds $\{a \geq 0\, :\, (a, b)\in H(P,Q) \}\subseteq H_b$. Furthermore, for $b\in H(Q)$,  the inequality is, indeed, an equality and it holds $\{a \geq 0\, :\, (a, b)\in H(P,Q) \}\supseteq H(P)$. This is not true outside $H(Q)$.

In both cases the diagonal line is on the points where $a=2g-b$, so that one can visually check the third item.
    \end{example}

    \graphhsmallone

    \graphntsmallone

Now, let $D=P_1+ \cdots +P_n$ be the sum of $n$ distinct
rational places different than $P$ and $Q$. 
For $b$ in ${\mathbb N}$ define 
$$H_b^*=\{a\geq 0 \, :\, C_\mathcal L(D, aP+bQ)\neq C_\mathcal L(D, (a-1)P+bQ) \}.$$
Notice that $H_b^*$ is always a finite set with $\# H_b^*\leq n$.
The next lemma relates the sets $H_b$ and $H_b^*$ at a given $b\in{\mathbb N}$.

\begin{lemma}\label{characterizationhbstar}
For $ b \in {\mathbb N}$ we have
\begin{enumerate}[(1)]
\item $H_b^* \subseteq H_b$, and 
\item $H_b \setminus H_b^*=\{ a \geq 1 \, :\, \ell(aP+bQ-D) \neq \ell((a-1)P+bQ-D)\}.$
\item 
If $0\leq a < n-b$ then $a \in H_b^*$ if and only if $a \in H_b$. 
\end{enumerate}
\end{lemma}

    \graphhsmalltwo   

\graphntsmalltwo

\begin{proof}
For any divisor $G$ with $\supp(G) \cap \supp(D)=\emptyset$, the evaluation map $ev : \cL(G) \rightarrow \fq^n$ has 
image the code $C_\mathcal L(D, G)$ and
kernel $\cL(G-D)$. This yields
\begin{align*}\label{dim}
&\dim C_\mathcal L(D, aP+bQ)-\dim C_\mathcal L(D, (a-1)P+bQ)=\\
&\ell(aP+bQ)-\ell((a-1)P+bQ)-\ell(aP+bQ-D)\\
&+\ell((a-1)P+bQ-D),
\end{align*} 
and items (1) and (2) follow.

Let us prove now item (3).
  If $0\leq a< n-b$ we have $0\leq a+b < n$.
  Applying Proposition~\ref{dimcode} yields $\dim C_\mathcal L(D, aP+bQ)=\ell(aP+bQ)$ and $\dim C_\mathcal L(D, (a-1)P+bQ)=\ell((a-1)P+bQ)$.
\end{proof}

    \begin{example}
Consider the Hermitian function field over ${\mathbb F}_4$. In Figure~\ref{fig:Hermitestarsetstwo} we depicted the sets $H_b$, and $H_b^*$,
while in Figure~\ref{fig:NormTracestarsetstwo}
      we depicted the same sets for the norm-trace function field over ${\mathbb F}_8$. The first and third items in 
Lemma~\ref{characterizationhbstar} can be checked.
That is, for each $b$ in both graphs, the set $H_b^*$ is included in $H_b$ and
below the lower diagonal line both sets coincide.
    \end{example}

The next lemma bounds the maximum element in $H_b^*$.

\begin{lemma}
  \label{hbstarbound}
Suppose that the genus of ${\cF}$ is nonzero.
  For $b\in{\mathbb N}$,
\begin{enumerate}[(1)]

\item $\max(H_b^*)\leq n+2g-1-b$. In particular, if $b >n+2g-1$ we have $H_b^* = \emptyset$.
\item $\max(H_b^*) \geq n-b$.
\end{enumerate}
\end{lemma}

\begin{proof}
\begin{enumerate}[(1)]

\item   By the Riemman-Roch Theorem,
  $\ell((n+2g-1-b)P+bQ-D)=g$ because $\deg((n+2g-1-b)P+bQ-D)=2g-1$,
  while $\ell((n+2g-1-b)P+bQ)=g+n$ because
  $\deg((n+2g-1-b)P+bQ-D)=2g-1+n\geq 2g-1$. On the other hand, by
  Proposition~\ref{dimcode}, $\dim C_\mathcal{L}(D,
  (n+2g-1-b)P+bQ)=\ell((n+2g-1-b)P+bQ)-\ell((n+2g-1-b)P+bQ-D)=n$.

  \item Let $a_0\geq 0$ be the minimum such that $\dim \cC_\cL(D, aP+bQ)=n,
  \, \forall a \geq a_0$. Since $\dim \cC_\cL(D, a_0P+bQ)
  =\ell(a_0P+bQ)-\ell(a_0P+bQ-D)$, we have that if $a_0< n-b-1$ then
  $\dim \cC_\cL(D,a_0P+bQ) =\ell(a_0P+bQ) \leq a_0+b +1<n$, a contradiction. We conclude that $a_0 \geq n-b-1$. If $\max(H_b^*) = n-b-1$ then $n=\dim \cC_\cL(D, (n-b-1)P+bQ)$ and as $\deg((n-b-1)P+bQ)=n-1<n$ then by Proposition \ref{dimcode} we have that $\ell((n-b-1)P+bQ)=n-g$, and we conclude $g=0$.
  \end{enumerate}
  \end{proof}

   \begin{example}
      In Figures~\ref{fig:Hermitestarsetstwo} 
      and \ref{fig:NormTracestarsetstwo}
      we also depicted the two diagonal lines representing the upper bound and the lower bound of Lemma~\ref{hbstarbound}.
It can be easily checked that for each $b$, the maximum of $H_b^*$ lies between the two lines.
    \end{example}

\section{The isometry-dual property for two-point codes}

The next definition was first introduced in \cite{GMRT}. 
\begin{definition}
A flag of codes $(C_i)_{i=0,\ldots, s}$ is said to satisfy the isometry-dual condition if there exist ${\bf x}\in (\mathbb{F}_q^*)^n$ such that $C_i$ is {\bf x}-isometric to $C_{s-i}^\bot$ for all $i=0,\ldots, s$. 
\end{definition}

In \cite{GMRT} the isometry-dual property was studied for one-point codes, that is, codes of the form $C_\cL(D, iP)$, where
$P\in\Pla(\cF)$ is a rational place, and $D$ is the sum of $n$ different rational places of $\Pla(\cF)$, all of them different than $P$.
The authors defined the set $H_D^*(P)=\{0\leq i \,: \, C_\cL(D, iP)\neq C_\cL(D, (i-1)P)\}$ and they proved that, given  $n>2g+2$,
the flag of one-point codes $C_\cL(D, iP)$, for $i \in H_D^*(P)$,
satisfies the isometry-dual condition if and only if the divisor $(n+2g-2)P-D$ is canonical
or, equivalently, $n+2g-1\in H_D^*(P)$. This result was extended for any $n \geq 2g+2$ in \cite{BDH}.
Next we extend this results to two-point codes.
Just notice that the two-point codes defined on a function field have length at most the number of rational places in the function field minus two, while the maximum length that one-point codes can have in the same function field is one more.

\begin{theorem}\label{t:isodual}
Let $\cF$  be a function field of genus $g$ over $\mathbb{F}_q$ and  $P, Q$ two different rational places in $\Pla(\cF)$.
Consider $b \in \N$ and the divisor $D=P_1+ \cdots +P_n$ the sum of $n\geq 2g+2b+2$ distinct
rational places, $P, Q \notin \supp(D)$. 

Let $H_b^\bullet=\{0\}\cup(H_b^*\cap\{1,\dots,n+2g-2-2b\})$.
The following are equivalent.
\begin{enumerate}[(1)]
\item There exists a constant vector $\bf x$ such that the flag of codes 
\begin{equation*}
\begin{split}
C_\mathcal L(D, a_0P+bQ)&\subsetneq
C_\mathcal L(D, a_1P+bQ)\subsetneq\\ 
&\dots \subsetneq C_\mathcal L(D, a_sP+bQ)
\end{split}
\end{equation*} 
is $\bf x$-isometry-dual, 
where $0\leq a_0<a_1<\dots < a_{s}$ is the ordered sequence of elements in $H_b^\bullet$.
\item
The divisor $E=(n+2g-2-2b)P+2bQ-D$ is canonical.
\item
$n+2g-1-2b \in H_{2b}^*$.
\end{enumerate}
\end{theorem}
\begin{proof}
Let us first prove that (1) implies (2). By hypothesis there exists $a$ such that
$2g\leq a\leq n-2-2b$. Since $ a\geq 2g$, by Lemma~\ref{hinhb} we have $a
\in  H_b$, and since $a \leq n-2-2b\leq n-1-b$, from Lemma~\ref{characterizationhbstar}(3) 
we obtain $a\in H_b^\bullet$.
Define $a^\perp=n+2g-2-2b-a$. As before, since $a\leq n-2-2b$, we have
$a^\perp\geq 2g$ and hence $a^\perp\in H_b$. Furthermore, since $a\geq
2g$ we also have $a^\perp\leq n-2-2b$, and so $a^\perp \in H_b^\bullet$.

Notice that, since $2g\leq a+b, a^\perp+b<n$, it holds $\dim(C_{\mathcal 
  L}(D, aP+bQ))=a+b+1-g$ and $\dim(C_{\mathcal 
  L}(D, a^\perp P+bQ))=a^\perp+b+1-g=n-\dim(C_{\mathcal 
  L}(D, aP+bQ))$. So, $a$ and $a^\perp$ correspond to dual indices in the isometry-dual flag.
We know that $C_\cL(D, aP+bQ)^\perp$ is $C_\cL(D, D+W-aP-bQ)$,
where $W$ is a canonical divisor  
with $v_P(W)=-1$ for any $P$ in $\supp(D)$ (see \cite[Proposition 2.2.10]{Stichtenoth}).
Assuming the ${\bf x}$-isometry-dual property we deduce that
$D+W-aP-bQ\sim a^\perp P+bQ$, since $a \geq 2g$.
Then $W\sim (a+a^\perp)P+2bQ-D=(n+2g-2-2b)P+2bQ-D=E$.
Hence $E$ is canonical.

Now we prove that (2) implies (1).
Suppose that $E=(n+2g-2-2b)P+2bQ-D$ is a canonical
divisor. Let $W$ be a canonical divisor  with $v_P(W)=-1$ for any $P$ in $\supp(D)$ (see \cite[Proposition 2.2.10]{Stichtenoth}).
Then there is a rational function $f$ such that $E+(f)=W$. 
In particular, $f$ has neither poles, nor zeros in the support of $D$.
Let ${\bf x}=ev_D(f)$. Then, for any $a\in H_b^\bullet$,
let $a^\perp=n+2g-2-2b-a$
and $(a^\perp)^*=\max\{a \in H_b^\bullet: a\leq a^\perp\}$.
Notice that $(a^\perp)^*\in H_b^\bullet$.
We have 
\begin{align*}
&D+W-(aP+bQ)\\
&=D+E+(f)-(aP+bQ)\\
&=(n+2g-2-2b)P+2bQ+(f)-(aP+bQ)\\
&=(a^\perp P+bQ)+(f).
\end{align*}
Hence, 
\begin{align*}
C_\cL(D, aP+bQ)^\perp&={\bf x}\cdot C_\cL(D, a^\perp P+bQ)\\
&={\bf x}\cdot C_\cL(D, (a^\perp)^* P+bQ).
\end{align*}

With this we proved that, assuming (2), there exists a vector ${\bf x}$ such that the dual code of any code of the form $C_\cL(D, aP+bQ)$, where  $a\in H_b^\bullet$
is exactly ${\bf x}\cdot C_\cL(D, a' P+bQ)$
for some $a'\in H_b^\bullet$.
Hence, the flag in (1) satisfies the isometry-dual property.

Now we prove that (2) and (3) are equivalent. 
By Riemann-Roch Theorem we know that $\ell(E+P)=g$ and, consequently, $\ell(E)\leq g$. Since $\deg(E)=2g-2$, $E$ is canonical if and only if $\ell(E)=g$ (see \cite[Proposition 1.6.2]{Stichtenoth}).
Then $E$ is canonical if and only if $\ell(E)=\ell(E+P)$, that is, if and only if 
$$\ell((n+2g-2-2b)P+2bQ-D)=\ell((n+2g-1-2b)P+2bQ-D).$$ By Lemma~\ref{characterizationhbstar}(2) this is equivalent to $n+2g-1-2b\in H^*_{2b}$.
\end{proof}

  Notice that the condition $n+2g-1-2b \in H_{2b}^*$
  is equivalent to the maximum of $H_{2b}^*$ attaining the upper bound in Lemma~\ref{hbstarbound}. If $b=0$ we are in the case of one-point codes and the condition $n+2g-1  \in H_{0}^*$ is coherent with \cite[Theorem 2]{BDH}.

\section{The isometry-dual property for two-point Hermitian codes}
\label{s:H}

In this section we study the isometry-dual property for two-point codes arising from the Hermitian function field. The results here, which are interesting on their own, will serve as an example for what will be proved about the isometry-dual property for Kummer extensions.

Consider the Hermitian function field $\cH$ over $\fqs$ of genus $g=\frac{q(q-1)}{2}$ defined by the affine equation $y^{q+1}=x^q+x$. Let $P=P_\infty $ denote the only pole of $x$ and $y$ in $\cH$. The Hermitian function field is  maximal over $\fqs$ with $q^3+1$ rational places denoted by 
$$\{ P_{\alpha,\beta} \, :\,   \beta^{q+1}=\alpha^q+\alpha, \alpha, \beta \in \fqs\}.$$
The study of a two-point code over the Hermitian function field does not depend on the fixed places $P$ and $Q$ since the automorphism group $Aut(\cH)$ of the Hermitian function field acts double transitively on the $\fqs$ rational places. An explicit proof of this fact can be found in \cite[Lemma 3.1]{HommaKim:Toward}. 

From now on we fix  $P=P_\infty $ and $Q=P_{0, 0}$ and consider the divisor 
$$D=\left(\sum_{\beta^{q+1}=\alpha^q+\alpha} P_{\alpha,\beta}\right)-Q$$ in order to study the family of algebraic geometry codes $C_{\cL}(D, aP+bQ)$ of length $n=q^3-1$.

The Riemann-Roch spaces $\cL(aP+bQ)$ and $\cL((a+q+1)P+(b-q-1)Q)$ are isometric by multiplication by the function $x$. 
By recursion,
suppose that $\theta, \rho$ are the quotient and remainder of the division of $b$ by $q+1$, that is, $b=\theta(q+1)+\rho$, $0\leq \rho\leq q, 0\leq \theta$. Then, it follows by induction on $\theta \geq 1$, that
\begin{center}
$a\in H_b^*$ if and only if $a+\theta(q+1)\in H_\rho^*$.
\end{center}

A complete characterization of the set $H_b^*$ can be found in \cite[Section 3]{HommaKim:Toward}, where the dimensions of the codes $C_{\cL}(D, aP+bQ)$ have been determined by Homma and Kim. Let $0\leq b\leq q$, then
\begin{equation*}\label{Hb*Hermitian}
\begin{split}
H_b^*&=\{ q\theta+\rho\,:\, 0\leq \rho \leq q-2, \, \rho\leq \theta \leq \rho+q^2-1\} \\
&\cup \{ q\theta +(q-1) \,:\,  \, q-b-1\leq \theta \leq q^2+q-b-3\}.
\end{split}
\end{equation*}
This characterization can be also found in the introductions of the papers 
\cite{HommaKim:Thecomplete} and \cite{HommaKim:Thetwopointcodes}.
We now compute the maximal element in $H_b^*$. 

From Equation \eqref{Hb*Hermitian} we have
\begin{align*}
\max(H_b^*) &=\max(H_\rho^*) -\theta(q+1)\\
&=
\begin{cases}
n+2g-(\theta+1)(q+1)  &\text{ if } 1\leq \rho \leq m-1,\\
n+2g-q-\theta(q+1)    &\text{ if }  \rho =0.
\end{cases}
\end{align*}
Now we can state the main theorem of this section.
\begin{theorem}\label{t:isodualseqH}
Let $n=q^3-1$ and suppose $b\leq n/2-g-1$.
Let $0 \leq a_0<\dots < a_{s}$ be the ordered sequence of elements in $H_b^\bullet=\{0\}\cup(H_b^*\cap\{1,\dots, n+2g-2-2b\})$.
The code flag 
\begin{equation*}
\begin{split}
C_\mathcal L(D, a_0P+bQ)&\subsetneq
C_\mathcal L(D, a_1P+bQ)\subsetneq\\ 
&\dots \subsetneq C_\mathcal L(D, a_sP+bQ)
\end{split}
\end{equation*} 
satisfies the isometry-dual condition
if and only if $q$ is even and $b$ is congruent with $q/2$ modulo $q+1$.
In particular, for odd $q$'s the flag of two-point codes  over the Hermitian function field is not isometry-dual for any $b$.
\end{theorem}
\begin{proof}
From Theorem \ref{t:isodual} and the description of the maximum of $H_b^*$ we deduce that the flag satisfies the isometry-dual condition
if and only if $2b$ is congruent with $q$ modulo $q+1$. 
\end{proof}

\begin{example}
  In Figure~\ref{fig:Hermiteisometrydualfour}
  there is a graph of the sets $H_b^*$
  for the two-point codes 
  $C_{\cL}(D, aP+bQ)$ defined on the Hermitian function field over ${\mathbb F}_{16}$.
  The sequences of places corresponding to codes satisfying the isometry-dual property are marked with black bullets.
  One can check all results analyzed up to this point for two-point Hermitian codes.
\end{example}

\graphh

\section{The isometry-dual property for two-point codes defined on Kummer extensions}
\label{s:K}

 Let $\cF/\fq$ be a  Kummer extension defined by the affine equation 
 \begin{equation}\label{Kummer}
 y^m=f(x)=\prod\limits_{i=1}^{r}(x-\alpha_i), \quad \alpha_i \in \fq
 \end{equation}
 where $f(x)$ is a separable polynomial of degree $2\leq r\leq m-1$ with $\gcd(m, r)=1$. This function field has genus $g=(m-1)(r-1)/2$ and only one rational place at infinity denoted by $P_\infty$, which is also the only pole of $x$ and $y$. Let $R_\infty \in \Pla(\fq(x))$ be the only pole of $x$, and $R_1, \dots, R_r$ stand for the places of the rational function field $\fq(x)$ associated to the zeros $\alpha_1,\ldots,\alpha_r$ of $f(x)$,  respectively. Since the places $R_1, \dots, R_r$ are totally ramified in the extension $\cF/\fq(x)$, there exists a unique rational place $P_{\alpha_i,0}$ in $\Pla(\cF)$ over $R_i$ for $i=1, \dots ,r$.
   Notice that $P_\infty$ in $\Pla(\cF)$ is the only place over $R_\infty$.
Then we have the following divisors in $\cF$:
\begin{enumerate}[(1)]
\item $(x-\alpha_i)=mP_{\alpha_i,0}-mP_\infty$  for every $i$, $1\le i \le r$,
\item $(y)= P_{\alpha_1,0}+\cdots +  P_{\alpha_r,0}-r P_\infty$.
\end{enumerate}

From now on in this section,
fix  $P=P_\infty $ and $Q=P_{\alpha_k,0}$ for some $k=1,\dots, r$.
Let $D$ be the sum of $n$ rational places in $\Pla(\cF)$ such that $P, Q \not\in \supp(D)$. We investigate the dimension of the two-point codes $C_\cL(D, aP+bQ)$ where $a\geq 0$ and $ b\geq 0$. First of all, notice that multiplication by the function $x-\alpha_k$ gives an isomorphism between the  Riemann-Roch spaces $\cL( aP+bQ)$ and  $\cL((a+m)P+(b-m)Q)$, and so the codes $C_\cL(D, aP+bQ)$ and $C_\cL(D, (a+m)P+(b-m)Q)$ are isometric.
In particular, if $b\geq m$, then $a\in H_b^*$ if and only if $a+m\in H_{b-m}^*$. Inductively, 
suppose that $\theta$ and $ \rho$ are the quotient and remainder of the division of $b$ by $m$, 
that is, $b=\theta m+\rho$, $0\leq \rho<m, 0\leq \theta$. Then,
\begin{center}
$a\in H_b^*$ if and only if $a+\theta m\in H_\rho^*$.
\end{center}

Next theorem is the principal result in this section. The notation $\alpha \mbox{ mod }m$ is used to refer to the unique integer congruent with $\alpha$ modulo $m$ in the interval from $0$ to $m-1$.

\begin{theorem}\label{t:kummer}
Let $\mathcal{F}/\fq$ be a Kummer extension of genus $g$  given by $$y^m=f(x)=\prod\limits_{i=1}^{r}(x-\alpha_i), $$  where $\alpha_i \in \fq ,$  $f(x)$ is a separable polynomial of degree $2\leq r \leq m-1$, and $\gcd(m, r)=1$. 
For $r_1 \leq q-r$, fix $R_{r+1}, \dots ,R_{r_1}$ rational places in $\Pla(\fq(x))$ completely split in the extension $\cF/\fq(x)$.
Let $P=P_\infty $ and $Q=P_{\alpha_k,0}$ for some $k=1,\dots, r$.
Consider the divisor 
$$D=\left(\sum\limits_{i=1}^{r} P_{\alpha_i,0}\right)-Q + \sum\limits_{i=r+1}^{r_1} \sum\limits_{\tilde P\mid R_i} \tilde P.$$ Denote $n=\deg(D)$ and suppose $n>2g-1.$

Let $1\leq \tilde r \leq m-1$ be the multiplicative inverse of $r$ modulo $m$. Then, $a\in H_b^*$ if and only if
$$
\begin{cases}
\tilde r a \mbox{ mod } m \leq \frac{a+b}{r-1}\text{, if }0\leq a+b<n,\\
\tilde r (a-n-1) \mbox{ mod } m > \frac{a+b-n}{r-1}\text{, if }n\leq a+b \leq n+2g-1.\\
\end{cases}
$$
\end{theorem}
\begin{proof}
  We use the same notations as above.   
  We have $n=r-1+(r_1-r)m\geq (m-1)(r-1)-1$. By Proposition \ref{dimcode}, $a \in H_b^*$ if and only if 
\begin{align*}
1&=\ell(aP+bQ)- \ell((a-1)P+bQ)\\
&-\ell(aP+bQ-D)+\ell((a-1)P+bQ-D).
\end{align*}
Hence, $a \in H_b^*$ if and only if 
\begin{equation}\label{dim1dim0}
\begin{split}
&\ell(aP+bQ)- \ell((a-1)P+bQ)=1 \, \mbox{and} \\
&\ell(aP+bQ-D)-\ell((a-1)P+bQ-D)=0.
\end{split}
\end{equation}
In the same way $R_1, \dots , R_r$ denote the places in the rational function field $\fq(x)$ associated to the zeros $\alpha_1,\ldots, \alpha_r$ of $f(x)$, we suppose the rational places $R_{r+1}, \dots ,R_{r_1}$ in $\Pla(\fq(x))$ are associated to the zeros of the functions $x- \alpha_{r+1}, \ldots, x- \alpha_{r_1} \in \fq(x)$, respectively. Then the divisor of the function $z=y\prod\limits_{i=r+1}^{r_1}(x-\alpha_i)$ in $\cF$ is 
$$(z)=D+Q-((r-1)+m(r_1-r)+1)P_\infty=D+Q-(n+1)P.$$
This yields the following equivalence of divisors 
$$D\sim (n+1)P-Q,$$
which allows to conclude 
\begin{equation}\label{equivdivisor}
\ell(aP+bQ-D)=\ell((a-n-1)P+(b+1)Q).
\end{equation}	
By \cite[Theorem 2.2]{Maharaj}
we have that for any $a\geq -1$ and $ b\geq 0$
\begin{align*}
\cL(aP+bQ)=\bigoplus\limits_{t=0}^{m-1}  \mathcal{L}([aP+bQ+(y^t)]_{|\fq(x)})\,y^t
\end{align*} 
which gives 
\begin{align*}
\ell(aP+bQ)&=\sum \limits_{t=0}^{m-1}  \ell([aP+bQ+(y^t)]_{|\fq(x)})\\
&=\sum \limits_{t=0}^{m-1}  \ell([aP+bQ+\sum\limits_{i=1}^{r} t P_i-rt P]_{|\fq(x)})\\
&=\sum \limits_{t=0}^{m-1}  \ell\left(\left\lfloor \frac{a-rt}{m} \right\rfloor R_\infty +\left\lfloor \frac{b+t}{m}  \right\rfloor R_k\right).\\
\end{align*} 
For $t=0, \ldots , m-1$ define $D_{t, a, b}=\left\lfloor \frac{a-rt}{m} \right\rfloor R_\infty +\left\lfloor \frac{b+t}{m}  \right\rfloor R_k.$
Then, using equation \ref{equivdivisor} we have
\begin{align*}
&\ell(aP+bQ)- \ell((a-1)P+bQ)= \\
&\sum \limits_{t=0}^{m-1} \left[\ell(D_{t, a, b})- \ell(D_{t, a-1, b) }\right], \\
&\text{ and }\\
&\ell(aP+bQ-D)-\ell((a-1)P+bQ-D)\\
&=\sum \limits_{t=0}^{m-1} \left[\ell( D_{t, a-n-1, b+1})- \ell( D_{t, a-n-2, b+1} )\right].
\end{align*} 
From equation \eqref{dim1dim0}, we have $a \in H_b^*$ if and only if 
\begin{align*}
&\sum \limits_{t=0}^{m-1} \left[\ell(D_{t, a, b})- \ell(D_{t, a-1, b) }\right]=1 \, \text{ and  }\\
&
\sum \limits_{t=0}^{m-1} \left[\ell( D_{t, a-n-1, b+1})- \ell(D_{t, a-n-2, b+1} )\right]=0.
\end{align*}
In the rational function field $\fq(x)$,  by the Riemann-Roch theorem,  we have that for any divisor $A$ with $\deg(A) \geq -1$, $\ell(A)=\deg(A)+1$. This yields two cases to be analyzed depending on the value of the sum $a+b$. 

Case 1: if $0\leq a+b<n$ then $\ell(aP+bQ-D)=\ell((a-1)P+bQ-D)=0$, and we are left to analyze the condition $\ell(aP+bQ)- \ell((a-1)P+bQ)=1$. Now we have
$\sum \limits_{t=0}^{m-1} \left[\ell(D_{t, a, b})- \ell(D_{t, a-1, b} )\right]=1$ if and only if the following two conditions are satisfied
\begin{equation}
\begin{split}
& \exists\, 0 \leq t_0 \leq m-1 \text{ such that } \deg(D_{t_0, a-1, b})  \geq -1  \text{ and  }\\
&\left\lfloor \frac{a-rt_0}{m} \right\rfloor=\left\lfloor \frac{a-1-rt_0}{m} \right\rfloor+1.
\end{split}
\end{equation}

Case 2: if $a+b\geq n>2g-1$, then by Riemann-Roch theorem
we have $\ell(aP+bQ)- \ell((a-1)P+bQ)=1$, and we are left to analyze the condition $\ell(aP+bQ-D)-\ell((a-1)P+bQ-D)=0$. In this case we get
$\sum \limits_{t=0}^{m-1} \left[\ell( D_{t, a-n-1, b+1})- \ell(D_{t, a-n-2, b+1} )\right]=0$ if and only if for every $t \in \{0, \dots, m-1\}$ one of the following conditions is satisfied:
\begin{enumerate}[i)]
\item $\deg(D_{t, a-n-1, b+1})=\left\lfloor \frac{a-n-1-rt}{m} \right\rfloor +\left\lfloor \frac{b+1+t}{m}  \right\rfloor <0$ or \\
\item $\deg(D_{t, a-n-1, b+1})=\left\lfloor \frac{a-n-1-rt}{m} \right\rfloor +\left\lfloor \frac{b+1+t}{m}  \right\rfloor  \geq 0$  and $ \left\lfloor \frac{a-n-1-rt}{m} \right\rfloor=\left\lfloor \frac{a-n-2-rt}{m} \right\rfloor.$
\end{enumerate} 
These conditions can be summarized as  $a \in H_b^*$  if and only if 
\begin{enumerate}[i)]
\item $\left\lfloor \frac{a+b-(r-1)t_0}{m}  \right\rfloor  \geq 0$  and $m \mid (a-rt_0)$ for some $0\leq t_0\leq m-1$  , if $0 \leq a+b<n$\\
\item $m\nmid (a-n-1-rt),$  for all $ t\in \left\{  0,\ldots,\min \left\{m-1,\lfloor \frac{a+b-n}{r-1}\rfloor \right\} \right\} $ , if $n\leq a+b\leq n+2g-1.$
\end{enumerate} 
Once more, we have $a \in H_b^*$  if and only if 
\begin{enumerate}[i)]
\item $ a \equiv rt_0 \mod{ m }$ for some $0\leq t_0\leq \min\{m-1,\lfloor\frac{a+b}{r-1}\rfloor\} $ if $0\leq a+b<n$.\\ 
 \item $a-n-1 \not\equiv rt \mod{ m },$ for all $t\in\left\{0,\ldots,\min\{m-1,\lfloor\frac{a+b-n}{r-1}\rfloor\}\right\} $ if $n\leq a+b\leq n+2g-1$. 
\end{enumerate} 
Now, since $\tilde r$ is the inverse of $r$ modulo $m$,
we deduce that $a \in H_b^*$  if and only if 
\begin{enumerate}[i)]
\item $\tilde ra\equiv t_0 \mod{ m }  \text{ for some }0\leq t_0\leq \min\{m-1,\lfloor\frac{a+b}{r-1}\rfloor\} \text{if } 0\leq a+b<n.$\\ 
\item $\tilde r(a-n-1)\not\equiv t \mod{ m } \text{ for all }t\in\left\{0,\ldots,\min\{m-1,\lfloor\frac{a+b-n}{r-1}\rfloor\}\right\}  \text{if } n\leq a+b\leq n+2g-1. $
\end{enumerate} 
And the result in the statement of the theorem follows.
\end{proof}

Notice that the divisor $D$ in Theorem \ref{t:kummer} has degree 
$n=r-1+M$, where
$M$ is a multiple of $m$ and
$n$ is assumed to be larger than $2g-1$. 
Of course, $n$ is also at most the total number of rational places on $\cF$ minus 2. 

The results of Theorem \ref{t:kummer} allow to compute the maximum of $H_b^*$ for Kummer extensions as follows in the next proposition.

\begin{proposition}\label{c:max}With hypotheses and notation as in Theorem~\ref{t:kummer}, and $b=\theta m + \rho$ where $0\leq \rho\leq m-1$, 
$$
\max(H_b^*)=\begin{cases}
n+2g-b +\rho - r-r\rho & \mbox{if } 0\leq \rho<\lfloor\frac{m}{r}\rfloor \\
n+2g-b+\rho -m & \mbox{if } \lfloor\frac{m}{r}\rfloor\leq  \rho\leq m-1.
\end{cases}
$$
\end{proposition}
\begin{proof}  
Since $\gcd(m, r)=1$, we can choose $1\leq u \leq m-1$ and $1\leq \lambda \leq r-1$ such that $\lambda m=ur+1$. In particular we have that $1\leq \tilde r= m-u \leq m-1$ is the inverse of $r$ modulo $m$. Notice that $u\equiv m-\tilde r$ modulo $m$ and that $\lfloor\frac{m}{r}\rfloor=\lfloor\frac{u}{\lambda}\rfloor$.
We recall that $2g=(m-1)(r-1)$.

First let us show that $a=n+2g-b+\rho - r-r\rho\in H_b^*$. By Theorem \ref{t:kummer}, we need to prove that $\tilde r (a-n-1)$ modulo $m$ is larger than $\frac{a+b-n}{r-1}$. 
In this case we have  $\tilde r (2g-b+\rho -r-r\rho-1) \equiv m-\rho-2 \mod m$, which is already between $0$ and $m-1$ and which is larger than $$\frac{a+b-n}{r-1}=\frac{2g+\rho-r-r\rho}{r-1}=m-\rho-2-\frac{1}{r-1},$$ as desired.

Let us now prove that $a=n+2g-b+\rho -m\in H_b^*$. We need to prove that $\tilde r (a-n-1)$ modulo $m$ is larger than $\frac{a+b-n}{r-1}$. 
We have  $\tilde r (2g-b+\rho-m-1)\equiv m-1 \mod m$, which is obviously larger than
$\frac{a+b-n}{r-1}=m-1-\frac{m-\rho}{r-1}$, as desired.

Now suppose $0\leq \rho < \lfloor\frac{m}{r}\rfloor$ and let $a=n+2g-b+\rho - r-r\rho+\ell$ with $\ell>0$.
We are going to prove that $\tilde r (a-n-1)$ modulo $m$ is at most $\frac{a+b-n}{r-1}=m-\rho-2+\frac{\ell-1}{r-1}$. 
We have 
$\tilde r(2g-b+\rho-r-r\rho+\ell-1) \equiv  \tilde r\ell -\rho -2 \mod m$.
Let $\lambda\ell+1=tr+s$ with $0\leq s\leq r-1$. Define
$$\mu=\frac{u}{\lambda}\left(-s+1+\left\lceil \frac{s}{r} \right\rceil r\right)+\dfrac{t+\lceil \frac{s}{r} \rceil}{\lambda}-\rho-2\;.$$
Notice that $\mu\in \mathbb{Z}$ and that $\mu$ is congruent with $\tilde r\ell -\rho -2$ modulo $m$. We are going to prove that $0\leq \mu\leq m-1$
and that  $\mu\leq\frac{a+b-n}{r-1}=m-\rho-2+\frac{\ell-1}{r-1}$,
proving hence that $n+2g-b+\rho - r-r\rho+\ell$ does not belong to $H_b^*$.
In fact, if we prove the last inequality, then
the inequality $\mu\leq m-1$ is a consequence of it, together with the bound 
$\ell\leq (\rho+1)(r-1)$, derived
from Lemma~\ref{hbstarbound}.

If $s=0$, then $\mu=\frac{u+t}{\lambda}-\rho-2=\frac{m+\ell}{r}-\rho-2$. Since $\gcd(\lambda, r)=1$ from the equality $\lambda(m+\ell)=r(u+t)$ we conclude that $r$ divides $m+\ell$. Hence $\rho+1\leq \lfloor \frac{m}{r}\rfloor < \frac{m}{r}<\frac{m+\ell}{r}$ and therefore $\mu\geq 0$.
On the other hand, as $r\geq 2$, if $\ell<r$ we have $\frac{m+\ell}{r}<m$ and then $\mu\leq m-\rho-2+\frac{\ell-1}{r-1}$. Now, if $\ell\geq r$ then $\frac{\ell}{r}\leq \frac{\ell-1}{r-1}$. So, $\mu=\frac{m}{r}-\rho-2+\frac{\ell}{r}\leq m-\rho-2+\frac{\ell-1}{r-1}$.

If $s> 0$, we have that $\lceil \frac{s}{r} \rceil=1$ and $\mu=\frac{u}{\lambda}(-s+1+r)+\frac{t+1}{\lambda}-\rho-2$ is a decreasing function of $s$.
The lowest value of $\mu$ is when $s=r-1$.
In this case $\lambda(m+\ell)+1=r(u+t+1)$ and so 
\begin{align*}
\mu &\geq \frac{2u+t+1}{\lambda}-\rho-2\\
&= \frac{u}{\lambda}+\frac{\lambda(m+\ell)+1}{\lambda r}-\rho-2\\
&=\frac{u}{\lambda}+\frac{1}{\lambda r}+\frac{m+\ell}{r}-\rho-2,
\end{align*}
then $\mu\geq \frac{m}{r}+\frac{m+\ell}{r}-\rho-2\geq 0$, where the last equality is because $\rho+1\leq \lfloor \frac{m}{r}\rfloor\leq \frac{m}{r}$ and $\frac{m+\ell}{r}\geq 1$.

Taking $s=1$ we have $\mu=\frac{ur}{\lambda}+\frac{t+1}{\lambda}-\rho-2=m-\rho-2+\frac{t}{\lambda}$. From $s=1$ we also obtain $\lambda\ell=tr$, and since $\lambda$ and $r$ are coprime we conclude $r$ divides $\ell$, and in particular $\ell\geq r$. Therefore $\frac{t}{\lambda}=\frac{\ell}{r}\leq \frac{\ell-1}{r-1}$. Hence, $\mu\leq m-\rho-2+\frac{\ell-1}{r-1}$,

Secondly  suppose 
$\lfloor\frac{m}{r}\rfloor\leq \rho\leq m-1$ and let 
$a=n+2g-b+\rho -m+\ell$ with $\ell>0$.
We need to prove that $\tilde r (a-n-1)$ modulo $m$ is at most $\frac{a+b-n}{r-1}=m-1+\frac{\ell-m+\rho}{r-1}$. 
We have $\tilde r (2g-b+\rho-m-1+\ell) \equiv \tilde r(\ell-r) \equiv m-1-u\ell \mod m$.
Let $\tilde \ell$ be the residue of $\ell$ modulo $m$ and 
$\mu=m-1-\left(\tilde \ell-\frac{\lfloor\frac{\lambda \tilde \ell-1}{r}\rfloor}{\lambda} r\right)u+\frac{\lfloor\frac{\lambda \tilde\ell-1}{r}\rfloor}{\lambda}$.

Notice that since 
$1\leq \lambda\tilde \ell-\lfloor\frac{\lambda\tilde \ell-1}{r}\rfloor r\leq r$,
one has 
$$-1<-1+\frac{1}{\lambda}+\frac{\lfloor\frac{\lambda\tilde \ell -1}{r}\rfloor}{\lambda}\leq\mu\leq m-1-\frac{u}{\lambda}+\frac{\lfloor\frac{\lambda\tilde\ell-1}{r}\rfloor}{\lambda}\leq m-1,$$ and so $\mu$ is the residue of $m-1-u\ell$ modulo $m$.
It remains to check that $\mu\leq m-1+\frac{\ell-m+\rho}{r-1}$.

Dividing $\lambda\tilde{\ell}-1$ by $r$ we get $\lambda\tilde{\ell}-1=t'r+s', 0\leq s'\leq r-1$. Then
\begin{align*}
\mu & =  m-1-\frac{u}{\lambda}(\lambda\tilde{\ell}-t'r)+\frac{t'}{\lambda} \\
 & = m-1 + \frac{t'-us'-u}{\lambda} \\
 & = m-1 + \frac{(t'-us'-u)(r-1)}{\lambda(r-1)}\\
&=m-1+\frac{\lambda\tilde{\ell}-s'-1-ur(s'+1)-t'+us'+u}{\lambda(r-1)}\\
&=m-1+\frac{\lambda\tilde{\ell}-(ur+1)(s'+1)-t'+us'+u}{\lambda(r-1)}\\
 & = m-1 + \frac{\lambda\tilde{\ell} -\lambda m(s'+1)-t'+u(s'+1)}{\lambda(r-1)}\\
 & = m-1 + \frac{\tilde{\ell}-m-ms'}{r-1}+\frac{u(s'+1)-t'}{\lambda(r-1)}\\
 & = m-1 + \frac{\tilde{\ell}-m+\rho}{r-1}-\frac{ms'+\rho}{r-1}+\frac{u(s'+1)-t'}{\lambda(r-1)}\\
 & = m-1 + \frac{\tilde{\ell}-m+\rho}{r-1}-\frac{\lambda ms'+\lambda\rho-us'-u+t'}{\lambda(r-1)}\\
& \leq m-1 + \frac{\ell-m+\rho}{r-1} - \frac{u(rs'-s'-1)+\lambda\rho+t'+s'}{\lambda(r-1)}.
\end{align*}
Let $A=\frac{u(rs'-s'-1)+\lambda\rho+t'+s'}{\lambda(r-1)}$, we need to prove that $A\geq 0$.
If $s'>0$, since $r \geq 2$,  then $A\geq 0$.
If $s'=0$, then we get $\lambda(\tilde{\ell}-m)=r(t'-u)$ and $r$ divides $\tilde{\ell}-m$ since $\gcd(\lambda, r)=1$. Then, from $\rho\geq \lfloor \frac{m}{r}\rfloor$, we obtain
$$A=\frac{t'-u+\lambda\rho}{\lambda(r-1)}=\frac{\frac{t'-u}{\lambda}+\rho}{r-1}=\frac{\frac{\tilde \ell-m}{r}+\rho}{r-1}\geq \frac{\left\lceil-\frac{m}{r}\right\rceil+\rho}{r-1}\geq 0 .$$ 
\end{proof}

\begin{theorem}\label{t:isodualKummer}
With the same notation as in Theorem \ref{t:kummer}, suppose that $0\leq b\leq n/2-g-1$.
Let $0=a_0<\dots < a_{s}$ be the ordered set of elements in $H_b^\bullet=\{0\}\cup(H_b^*\cap\{1,\dots, n+2g-2-2b\})$.
The flag of algebraic geometry codes 
\begin{equation*}
\begin{split}
C_\mathcal L(D, a_0P+bQ)&\subsetneq
C_\mathcal L(D, a_1P+bQ)\subsetneq\\ 
&\dots \subsetneq C_\mathcal L(D, a_sP+bQ)
\end{split}
\end{equation*} 
is ${\bf x}$-isometry-dual for some vector ${\bf x}$
if and only if $m$ divides $2b+1$.
\end{theorem}
\begin{proof}
  By Theorem~\ref{t:isodual},
  the flag above is ${\bf x}$-isometry-dual for some vector ${\bf x}$ if and only if
  $n+2g-1-2b\in H_{2b}^*$.
  Now, by Theorem~\ref{t:kummer},
  $n+2g-1-2b\in H_{2b}^*$ if and only if
  the remainder of  $\tilde r ((n+2g-1-2b)-n-1)$ divided by $m$ is larger than $\frac{2g-1}{r-1}$.
  But, since $2g=(m-1)(r-1)$, this is equivalent to the remainder of
    $\tilde r (-r-2b-1)$ divided by $m$ being exactly $m-1$.
  That is, if and only if
  $-r-2b-1\equiv -r$ modulo $m$, i.e., $2b+1\equiv 0$ modulo $m$.
    \end{proof}

\begin{corollary}\label{c:noisodualformeven}
  If $m$ is even, there is no isometry-dual flag of codes
  for $0\leq b\leq n/2-g-1$.
 If $m$ is odd, there is an isometry-dual flag of codes at a given integer $b$,
  for $0\leq b\leq n/2-g-1$,
  if and only if $b=\frac{m\tilde t-1}{2}$ for some $\tilde t$, which is necessarily odd. Hence, $b=mt+\frac{m-1}{2}$ for some
  $0\leq t\leq \lfloor\frac{n-m-2g-1}{2m}\rfloor$.
  \end{corollary}

\begin{remark}
  With the same notation as in Theorem \ref{t:kummer}, let $n$ stand for the total number of rational places  on the function field. Notice that Theorem~\ref{t:isodualKummer}, in the case $b=0$, corresponds to flags of one-point algebraic geometry codes. In particular, it is shown that there is no isometry-dual flag of one-point codes of length $n-2$ for the case of the Kummer extensions we analyzed.

  This is coherent with the existence of isometry-dual flags of one-point codes of length $n-1$ such as the Hermitian function fields examples is 
\cite{GMRT} and \cite{BDH}.
Indeed, in \cite[Theorem 23]{BDH} it was shown that unless the Weierstrass semigroup at $P$ is ${\mathbb N}$, if a flag of codes of length $n-1$ satisfies the isometry-dual condition then the flag obtained by shortening it at one place (that is, of length $n-2$) does not satisfy the isometry-dual condition.
\end{remark}

In what follows we illustrate the results in Corollary \ref{c:max} and Theorem \ref{t:isodualKummer} for two function fields.

\begin{example}\label{e:kondo}
In this first example we work with a maximal function field that has been object of investigation in \cite{GV1986}. It is called the generalized Hermitian function field.
It is defined by the affine equation $y^{q^\ell+1}=x^q+x$ over $\mathbb{F}_{q^{2\ell}}$ with $\ell$ odd. It has genus $g = q^\ell(q-1)/2$, one single place at infinity $P_\infty = (1 : 0 : 0)$ and plus 
$q^{2\ell+1}$ rational places.

We take $P=P_\infty$ and $Q=P_{0,0}$ and consider the codes
$C_{\cL}(D, aP+bQ)$ of length $n=q^{2\ell+1}-1$.

If $m$ is even (i.e. $q$ odd), by Corollary~\ref{c:noisodualformeven}, there will be no isometry-dual flags for $0\leq b\leq n/2-g-1$.
So, we will analyze only the cases in which $m$ is odd, and so the characteristic is $2$.
In this case, Theorem~\ref{t:isodualKummer} and Corollary~\ref{c:noisodualformeven} ensure that
for $0 \leq b \leq n/2-g-1$, the flag of codes associated to $H_b^\bullet$ satisfies the isometry-dual property
if and only if $b=(q^\ell+1)t+\frac{q^\ell}{2}$ for $t=0, \dots, \lfloor\frac{(q^\ell-1)q^{\ell+1}-3}{2(q^\ell+1)}\rfloor$.

In Figure~\ref{fig:Kondoisometrydualtwo}  there is a graph of the sets $H_b^*$ for the two-point codes 
$C_{\cL}(D, aP+bQ)$, where $P=P_\infty$ and $Q$ is the place at the origin defined over ${\mathbb F}_{64}$, with $q=2$ and $\ell=3$.
The sequences of places corresponding to codes satisfying the isometry-dual property are marked with black bullets. As just proved, they correspond 
to the values of $b$ equal to $9t + 4$ for $t=0, \dots, 6$.
\end{example}
\graphk

\begin{example}\label{e:normtrace}
In this second example we work with the norm-trace function field that was first addressed by Geil in \cite{Geil}. In \cite{KimLee} there is a study of two-point codes over this function field. Let $\ell \geq 2$. The norm-trace function field is defined by the affine equation
$y^{\frac{q^\ell-1}{q-1}}=x^{q^{\ell-1}}+x^{q^{\ell-2}}+\dots +x$.
It has genus $g = \frac{(q^{\ell-1}-1)}{2} (\frac{q^\ell -1}{ q-1}-1)$,  and $q^{2\ell-1}+1$ rational places over $\mathbb{F}_{q^{\ell}}.$

We take $P=P_\infty$ and $Q$ at the origin and consider the codes
$C_{\cL}(D, aP+bQ)$ of length $n=q^{2\ell-1}-1$.

If $m$ is even (this occurs if and only if $q$ is odd and $\ell$ is even), by Corollary~\ref{c:noisodualformeven}, there will be no isometry-dual flags for $0\leq b\leq n/2-g-1$.
  So, we will analyze only the cases in which $m$ is odd.
In this case, Theorem~\ref{t:isodualKummer} and Corollary~\ref{c:noisodualformeven} ensure that
for $0 \leq b \leq n/2-g-1$, the flag of codes associated to $H_b^\bullet$ satisfies the isometry-dual property
if and only if 
\begin{align*}
&b=\frac{q^\ell-1}{q-1}t+\frac{\frac{q^\ell-1}{q-1}-1}{2}, \text{ for }\\
&t=0, \dots, \lfloor\frac{(q^{2\ell-1}-2)(q-1)-(q^{\ell}-1)-(q^{\ell-1}-1)(q^\ell-q)}{2(q^\ell-1)}\rfloor
\\&=\frac{q^{2\ell}-2q^{2\ell-1}+q^\ell-3q+3}{2(q^\ell-1)}.
\end{align*}
Notice that for $q=2$,
$t$ attains only one value and so there is only one isometry-dual flag of 
codes for a fixed $b\leq n/2-g-1$.

In Figure~\ref{fig:NormTraceisometrydualtwo}
  there is a graph of the sets $H_b^*$
for the norm-trace function field over ${\mathbb F}_{8}$, i.e, with $q=2$ and $\ell=3$.
  The sequences of places corresponding to codes satisfying the isometry-dual property are marked with black bullets.
  One can check all results analyzed up to this point for two-point norm-trace codes.
In this case,
    for $0 \leq b \leq n/2-g-1$, the flag of codes associated to $H_b^\bullet$ satisfies the isometry-dual property
    if and only if $b=7t+3$ for $t=0$.
    That is, if and only if $b=3$.  
\end{example}
\graphnt


\bibliographystyle{plain}

\begin{IEEEbiographynophoto}{Maria Bras-Amorós}
received her PhD in Applied Mathematics in 2003 from Universitat Politècnica de Catalunya and part of her doctoral and postdoctoral work was developed at San Diego State University, California. She is a full professor at Universitat Rovira i Virgili in Tarragona, Catalonia. She has been with Universitat Politècnica de Catalunya and Universitat Autònoma de Barcelona. Her main research interests are in the area of coding theory and discrete mathematics.
\end{IEEEbiographynophoto}

\begin{IEEEbiographynophoto}{Luciane Quoos} 
 received 
her M.Sc. and Ph.D. in mathematics from Instituto Nacional de Matem\'atica Pura e Aplicada (IMPA) in 1995 
and 2000, respectively, both in Brazil. In 1998 she joined the Instituto de Matem\'atica at the Universidade Federal do Rio de Janeiro where she is currently a Full Professor.  During a nine month period, starting in August 2002, she held an 
ERCIM (European Research 
Consortium for Informatics and 
Mathematics) postdoctoral fellowship at the Norwegian University of Science and Technology.
 In 2016 she spent her sabbatical at the Universit\`a degli Studi di Perugia, Italy.
Her research interests
include maximal curves and curves with many rational points over finite fields, 
Weierstrass semigroups, and algebraic geometric codes. 
\end{IEEEbiographynophoto}

\begin{IEEEbiographynophoto}{Alonso Sep\'ulveda Castellanos}
received a B.Sc. in mathematics from Universidad
Industrial de Santander, Colombia, in 2002, and an M.Sc. and Ph.D. in mathematics
from Universidade Estadual de
Campinas, Brazil, in 2004 and 2008, respectively. Since 2008, he is with the Faculdade de Matem\'atica, Universidade
Federal de Uberl\^andia, Brazil. His research interests lie in algebraic curves over Finite Fields and their applications to Coding Theory and Cryptography, Weierstrass and Numerical Semigroups, and applications in finite fields.
\end{IEEEbiographynophoto}

\end{document}